\documentclass[iop]{emulateapj}
\usepackage{bm}
\usepackage{multirow}
\usepackage{color}
\usepackage{float}
\usepackage{soul}
\usepackage{mathrsfs}
\usepackage{epsfig}
\usepackage{graphicx}
\usepackage{mathtools}
\usepackage{color}

\usepackage{natbib}
\usepackage{amsmath}
\usepackage{amsfonts}
\usepackage{amssymb}
\usepackage{epsfig}
\usepackage{hyperref}
\usepackage{graphicx}
\usepackage{booktabs}

\newcommand{\beq}{\begin{equation}}
\newcommand{\eeq}{\end{equation}}
\newcommand{\beqa}{\begin{eqnarray}}
\newcommand{\eeqa}{\end{eqnarray}}

\newcommand{\pc}{\rm{21cmFAST}}

\newcommand{\gs}{\rm{global 21cm signal}}

\begin{document}
\title{Enhanced global signal of neutral hydrogen due to excess radiation at cosmic dawn}

\author{Chang Feng\footnote{changf@illinois.edu}}
\affiliation{Department of Physics, University of Illinois at Urbana-Champaign, 1110 W Green St, Urbana, IL, 61801, USA}

\author{Gilbert Holder}
\affiliation{Department of Physics, University of Illinois at Urbana-Champaign, 1110 W Green St, Urbana, IL, 61801, USA}
\affiliation{Department of Astronomy, University of Illinois at Urbana-Champaign, 1002 W Green St, Urbana, IL, 61801, USA}
\affiliation{Canadian Institute for Advanced Research, Toronto, ON, Canada}

\begin{abstract}
We revisit the \gs\ calculation incorporating a possible radio background at early times, and find that the \gs\ shows a much stronger absorption feature, which could enhance detection prospects for future 21 cm experiments. In light of recent reports of a possible low-frequency excess radio background, we propose that detailed 21 cm calculations should include a possible early radio background.
\end{abstract}

\maketitle

\section{Introduction}

The redshifted 21 cm emission, due to the transition between the triplet and singlet states of neutral hydrogen, is an important probe of the high-redshift universe. Unlike the cosmic microwave background (CMB), the 21 cm emission maps the three dimensional distribution of neutral hydrogen. It contains both spatial and spectral information, leading to experimental approaches focused primarily on the spatial fluctuations or primarily on the sky-averaged spectrum. The latter is the so-called \gs, which is an important cosmological probe of the epoch of reionization (EoR) \citep{1999A&A...345..380S, 2006MNRAS.371..867F}.

The 21 cm spatial fluctuations have been extensively studied theoretically \citep{2004ApJ...613....1F,2010ARA&A..48..127M}, but this faint signal is yet to be detected, largely due to very bright foreground contaminants, such as Galactic synchrotron, supernovae remnants, free-free emission and radio point sources \citep{2006PhR...433..181F}. Many experiments, such as the Low-Frequency Array (LOFAR) \citep{2013A&A...556A...2V}, the Square Kilometre
Array (SKA) \citep{2015aska.confE...1K}, the Murchison Widefield Array (MWA) \citep{2009IEEEP..97.1497L} and the Hydrogen Epoch of
Reionization Array (HERA) \citep{2017PASP..129d5001D}, are designed to study the 21 cm signatures and foreground mitigation strategies have been proposed making use of the smooth spectral behavior of foregrounds.

The \gs\ may enable a first detection of signatures in the dark ages; only a small amount of integration time ($\sim\mathcal{O}(10^2)$ hours) is required to reach a sufficient detection sensitivity while interferometric arrays designed for 21 cm fluctuation measurements, such as LOFAR, would require significantly longer integration time. There are difficult calibration problems, but a single-dipole experiment can be constructed to probe the \gs. A few ongoing projects including the Dark Ages Radio Experiment (DARE) \citep{2012AdSpR..49..433B}, the Experiment to Detect the Reionization Step (EDGES) \citep{2010Natur.468..796B}, the Large-Aperture Experiment to Detect the Dark Ages (LEDA) \citep{2012arXiv1201.1700G} and the Long Wavelength Array (LWA) \citep{2010iska.meetE..24H}, may be able to detect the \gs.  However, substantial challenges still exist for the \gs\ measurements. Besides those are shared with the fluctuation measurements, the ionospheric reflection can generate a contaminating power that may be two to three orders of magnitude higher than the global 21 cm signal \citep{2014arXiv1409.0513D,2014MNRAS.437.1056V}. A mathematical formalism for isolating these foregrounds has been established \citep{2013PhRvD..87d3002L} and the separation of the \gs\ is aided by low resolution spatial information. A Bayesian inference technique was also devised to isolate the foregrounds from the raw signals \citep{2017ApJ...844...33B}. On the other hand, a space mission, such as DARE \citep{2012AdSpR..49..433B} could sidestep some of the issues such as atmospheric contaminant and terrestrial radio frequency interference. With current techniques, foregrounds such as Galactic synchrotron and free-free emission, while orders of magnitude brighter than the 21 cm signal, can be largely separated from the data and detection of the \gs\ can be achieved by these dedicated experiments. 

The characteristic peaks and troughs of the \gs\ are rich in astrophysical information, especially heating mechanisms and ionizing sources. It is believed that gas heating in the early universe mainly comes from accreting black holes and the first stars. The locations of the troughs can differentiate metal-free primeval stars from metal-rich \citep{2017ApJ...844...33B}, extra heating due to dark matter annihilation can reduce the absorption trough significantly \citep{2013MNRAS.429.1705V}, while warm dark matter, which suppresses structure formation, can delay the absorption feature in the \gs \citep{2014MNRAS.438.2664S}. 

Numerical simulations have shown that the redshift evolution of the power spectrum of the brightness spatial fluctuations at fixed angular scales resembles that of the \gs, although spatial fluctuations are roughly one order of magnitude fainter than the monopole \citep{2008ApJ...689....1S}. This means that fluctuation experiments with interferometric arrays can also study the global signatures.

The brightness temperature fluctuation is set by the difference between the spin temperature of the hydrogen gas and the temperature of the local radiation field at the 21 cm frequencies. It is usually assumed that the radiation field is dominated by the CMB. Recently, a balloon-borne double-nulled instrument called the Absolute Radiometer for Cosmology, Astrophysics and Diffuse Emission (ARCADE 2) detected excess radio radiation, revealing a strong radio radiation background, consistent with CMB radiation at high frequencies but significantly deviating from a black-body spectrum at low frequencies \citep{2011ApJ...734....5F}. This radiation is substantially larger than expected from observed radio counts \citep{2018PASP..130c6001S}. It is possible that some of this radiation could originate from early times, such as radio-loud quasars, which are recently studied \citep{2018arXiv180205880B}. Excess radiation at early times at radio wavelengths can change the evolution of the spin temperature and also provide an enhanced background against which absorption can be observed. This suggests that a correction to the spin temperature of neutral hydrogen may be required at high redshift and new features of the \gs\ are possible. If neglected, the predicted intensity of the \gs\ could be underestimated. In this work, we revisit the calculation of \gs\ with a possible high redshift radio radiation field and discuss the implications of the new features for the \gs. \\

\section{Observational radio radiation excess detected by ARCADE 2}
\begin{figure}
\includegraphics[width=10cm, height=8cm]{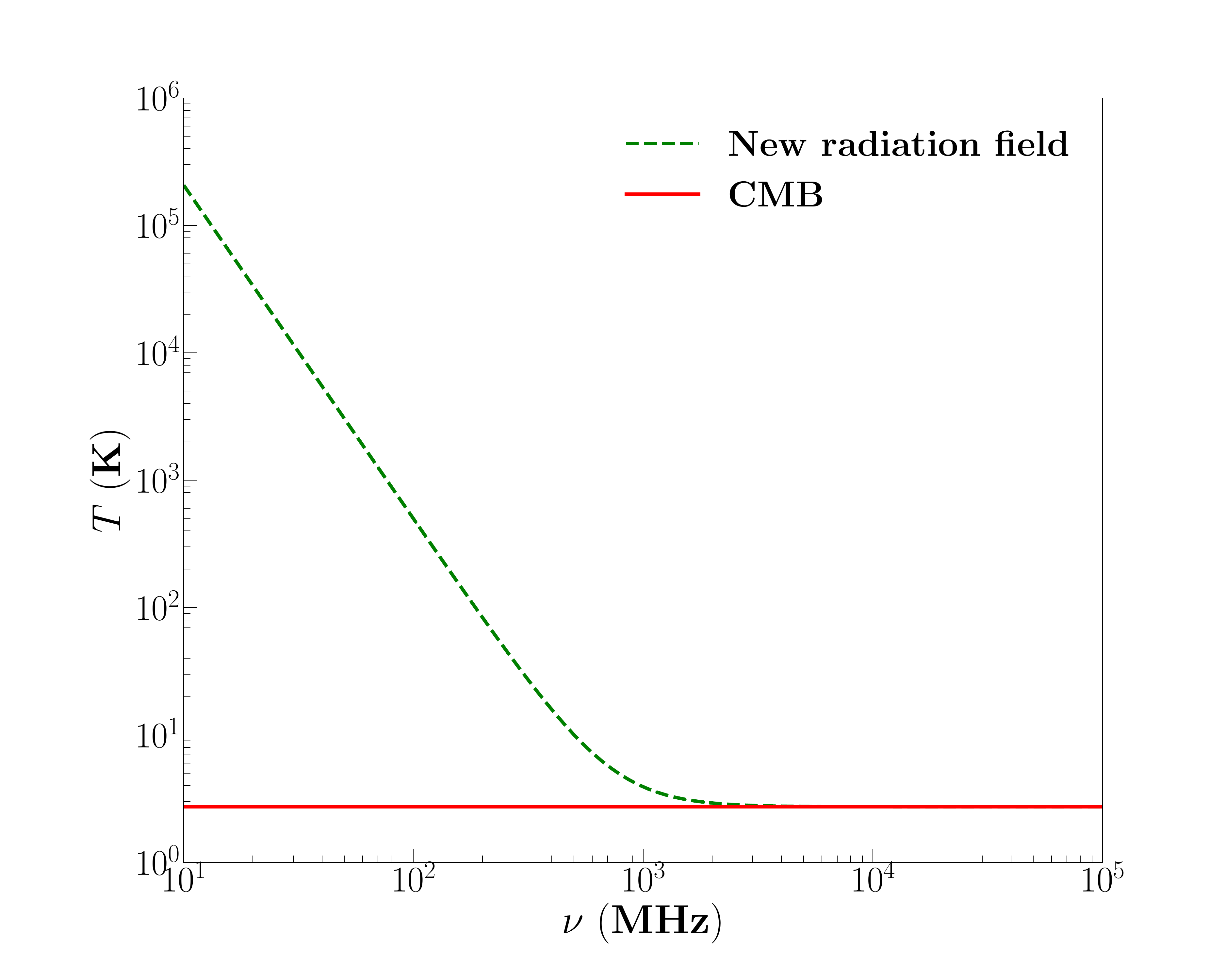}
\caption{The radiation excess detected by ARCADE 2. The red solid line represents a CMB radiation field, while the green dashed line shows a significant deviation at $\nu< 1\rm{GHz}$. These low frequencies are relevant to the signatures coming from high redshift.}\label{radexcess}
\end{figure}

The detected ARCADE 2 excess has been fitted by a simple power law, shown in Fig.~\ref{radexcess}. It is seen that the radiation is dominated by the CMB at high frequencies but there is a substantial radio background that dominates at lower frequencies. The radiation excess is fitted with a simple power law as
\begin{equation}
T(\nu)=T_{\rm CMB}+\xi T_R\Big(\frac{\nu}{\nu_0}\Big)^{\beta},
\end{equation}
where the CMB temperature is $T_{\rm CMB}$ = 2.729 K, normalized to $T_R$ = 1.19 K at a reference frequency $\nu_0$ = 1 GHz, with a spectral index $\beta$ = -2.62. The second term on the right hand side is the excess radiation \citep{2018PASP..130c6001S}. The excess fraction $\xi$ is set to be unity \citep{2011ApJ...734....5F} but given the fact that radio source counts can contribute to the background, we consider the excess fraction at early times as a free parameter. The excess can not be easily explained by Galactic emission, unresolved emission from known radio point source population or CMB spectral distortions. The flatter spectrum of the detected excess differentiates it from radio point sources which have a much deeper spectrum \citep{0004-637X-734-1-6}. Among other possibilities, some of this excess could also originate from high redshift. Because of the steeply rising spectrum toward longer wavelengths, a significant amount of this radiation that has been redshifted from early times would be substantially larger than the CMB contribution at a rest wavelength of 21 cm at high redshift.

\begin{figure}
\includegraphics[width=10cm, height=8cm]{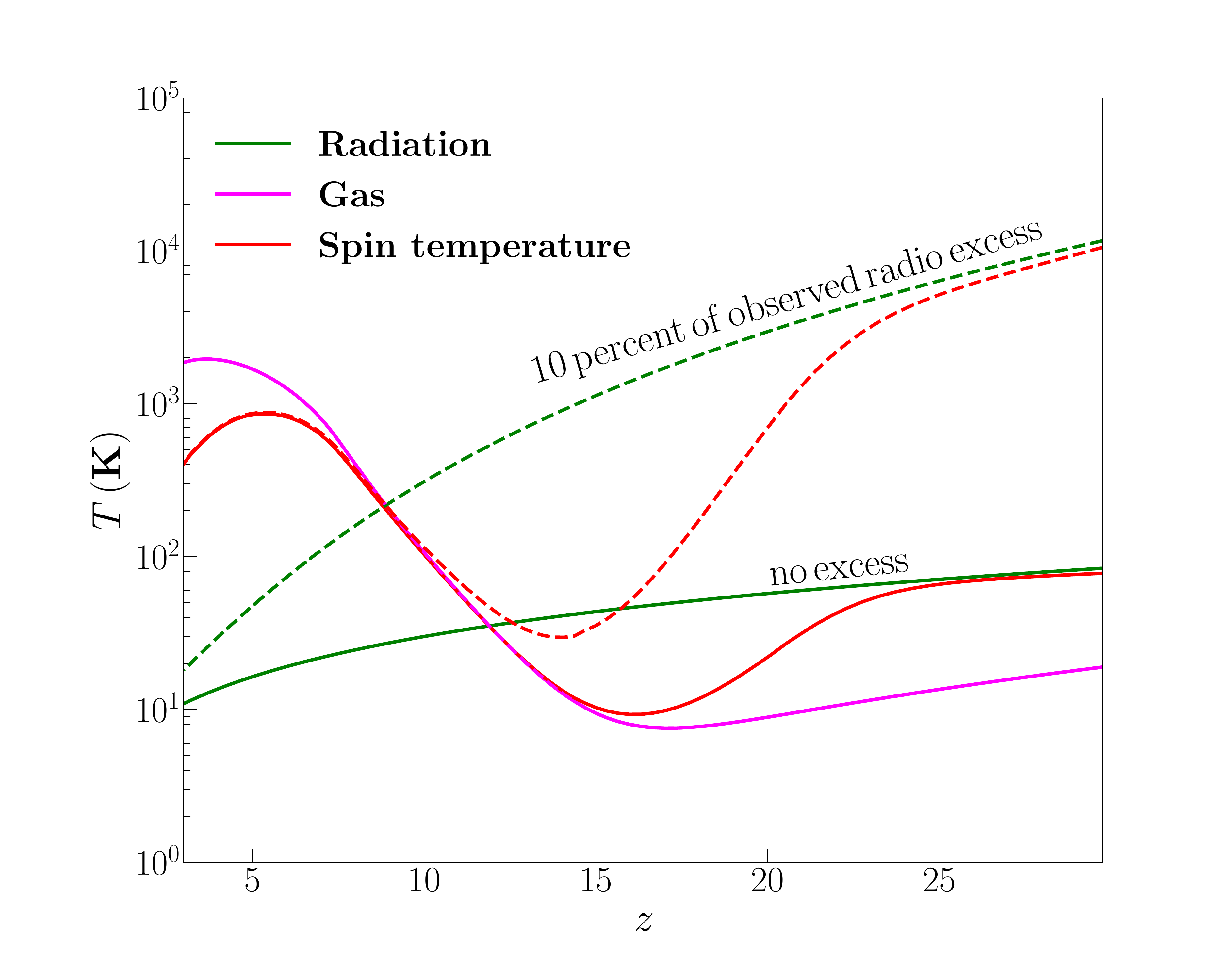}
\caption{The spin, gas and radiation temperatures. We use the default setting of the code \pc\ to calculate all of these temperatures. The temperatures are shown with no excess (solid lines) and with 10\% excess (dashed lines).}\label{gs_standard}
\end{figure}

\section{Modified 21 cm global signal}

In this section, we derive the calculations of the \gs\ in the presence of an excess radio radiation field. The global signal is modeled as
\begin{equation}
\delta T_b=27x_{\rm HI}\Big(1-\frac{T_r}{T_{\rm s}}\Big)\sqrt{\frac{1+z}{10}\frac{0.15}{\Omega_mh^2}}\Big(\frac{\Omega_bh^2}{0.023}\Big) ({\rm mK}).\label{gs}
\end{equation}
Here $x_{\rm HI}$ is the fraction of the neutral hydrogen, $T_r$ and $T_s$ are radiation and spin temperatures, respectively. We use the latest Planck parameters for the dark matter ($\Omega_m$) and baryonic matter density ($\Omega_b$) fractions \citep{2016A&A...594A..13P}.  Also, the variable $T_r$ is replaced by the fitted model in the last section and $\nu=\nu_{\rm HI}/(1+z)$ with $\nu_{\rm HI}=1420\,\rm{MHz}$.

The spin temperature $T_s$ is coupled to the gas temperature $T_K$ by collisions and by the Wouthuysen-Field effect (WFE) \citep{1952AJ.....57R..31W,1959ApJ...129..536F}. It is described as
\begin{equation}
T_s^{-1}=\frac{T_r^{-1}+x_{\alpha}T_{\alpha}^{-1}+x_cT_K^{-1}}{1+x_{\alpha}+x_c},\label{eqspin}
\end{equation}
where $T_{\alpha}$ is the color temperature of radiation around the Lyman-alpha transition, $x_{\alpha}$ and $x_c$ are WFE and collisional coupling coefficients, which both follow a simple scaling $\propto T_r^{-1}$ \citep{2012RPPh...75h6901P}. We run \pc\ \footnotemark[1]\footnotetext[1]{\url{https://github.com/andreimesinger/21cmFAST}} to create templates for the ionizing fraction $x_i$, $x_{\alpha}$, $x_c$ and $x_{\alpha}/T_{\alpha}$ \citep{2011MNRAS.411..955M}.  In Fig.~\ref{gs_standard}, we show detailed calculations of spin temperature, gas temperature and radiation temperature from redshift $z=4$ to 30 in solid and dashed lines, corresponding to no radiation excess and an excess corresponding to 10\% of the background today being in place at early times, respectively. The gas ($T_K$) temperature may depend on the CMB at early times through Thomson scattering but will not be strongly affected by the relatively small amount of energy in the non-thermal radio radiation fields. As an approximation, we assume both $T_{\alpha}$ and $T_K$ are independent of the excess radiation field. We substitute the radiation field $T_r$ in Eq. (\ref{eqspin}) by the fitted model of the excess in Fig.~\ref{radexcess}, in this case reducing the excess to only 10\% of the observed value to account for possible contributions from low redshift. The coupling coefficients $x_{\alpha}$ and $x_c$ are modified via a simple scaling $1/T_r$ and the new spin temperature in Eq. (\ref{eqspin}) is derived for this new radiation field.\\

\begin{figure}
\includegraphics[width=10cm, height=8cm]{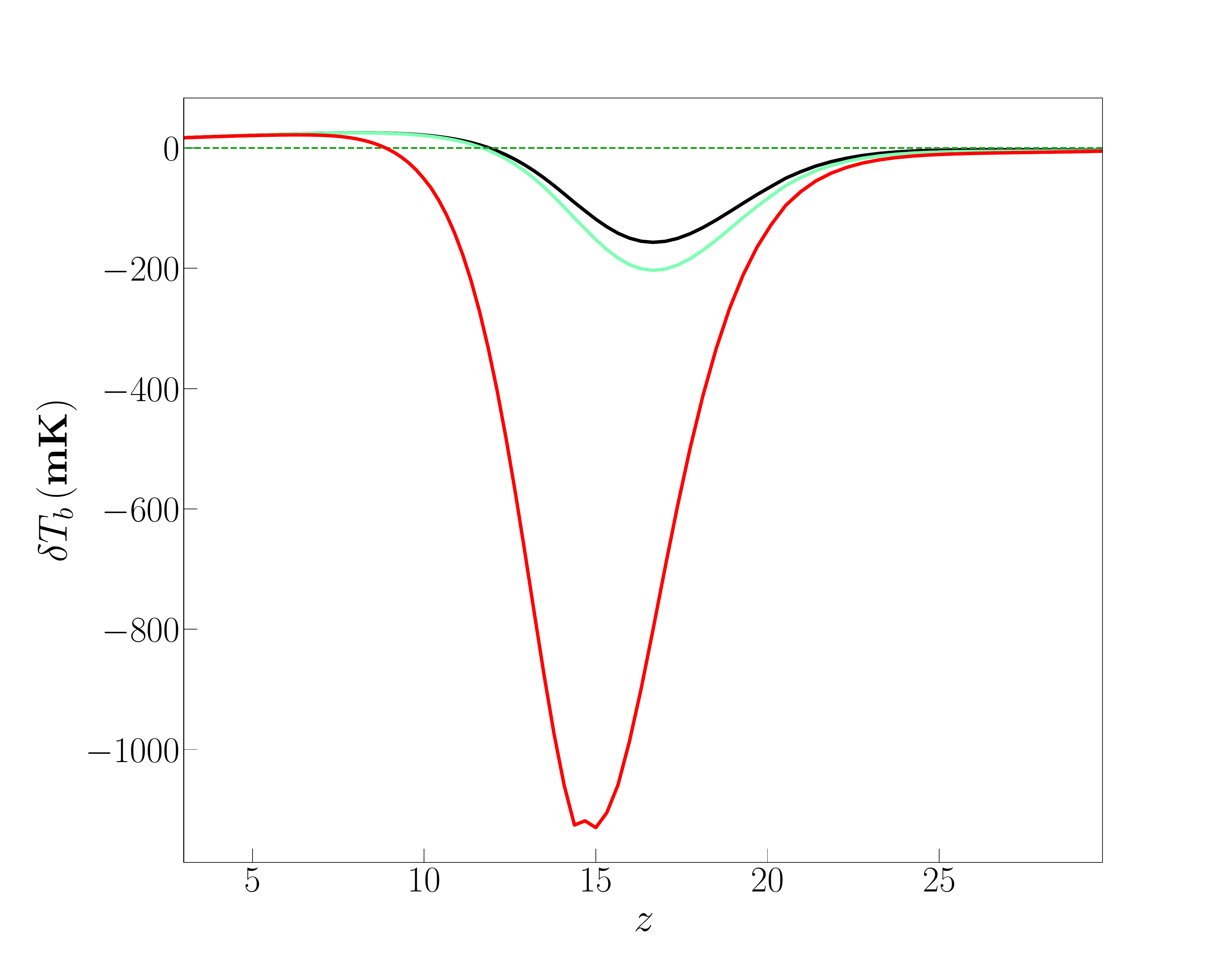}
\includegraphics[width=10cm, height=8cm]{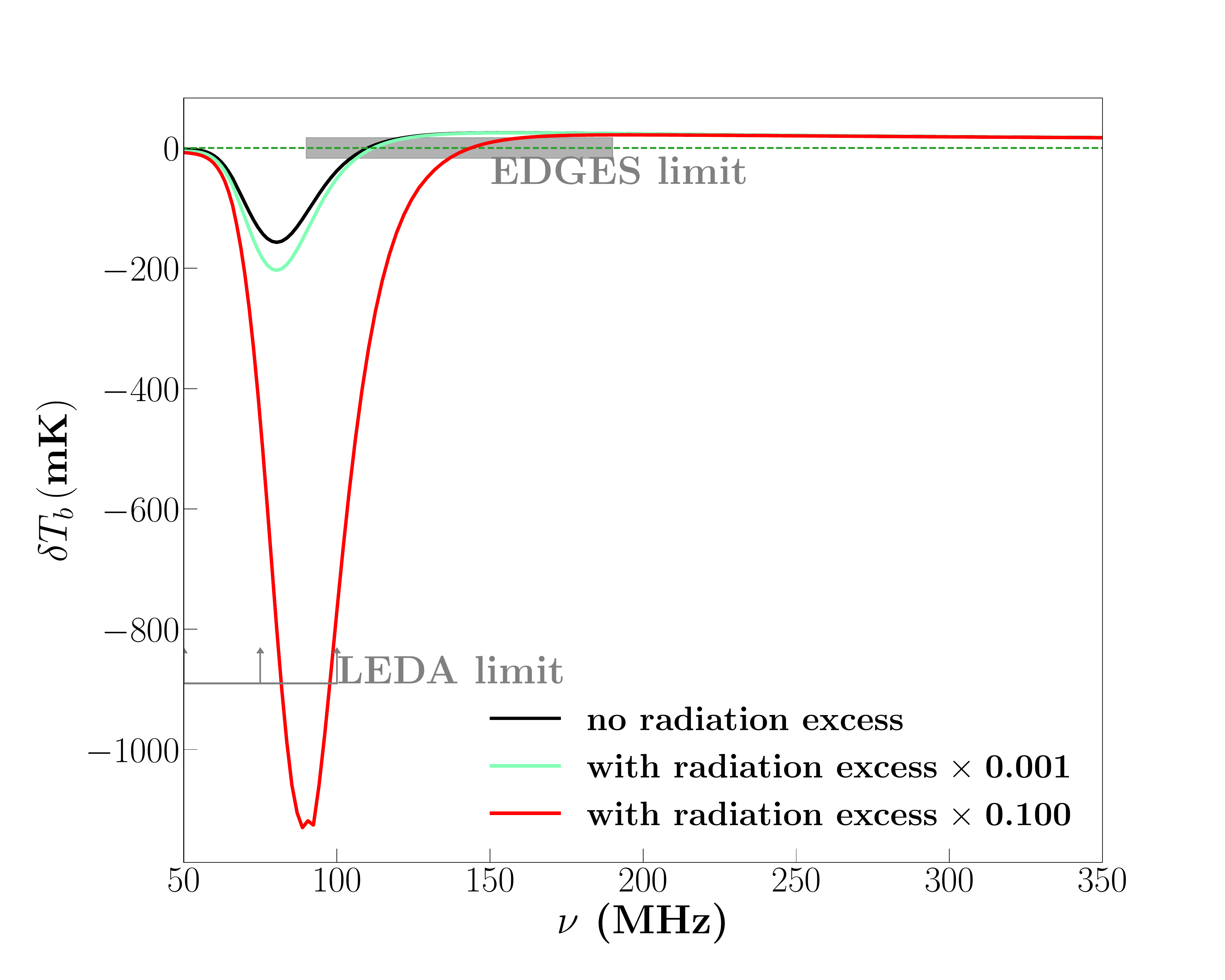}
\caption{The enhanced \gs\ as a function of redshift $z$ (top) or frequency $\nu$ (bottom). The trough located at $\nu\sim100\,\rm{MHz}$ will trace physical signatures in the cosmic dawn epoch at $z>12$.}\label{gs}
\end{figure}

\begin{figure}
\includegraphics[width=10cm, height=8cm]{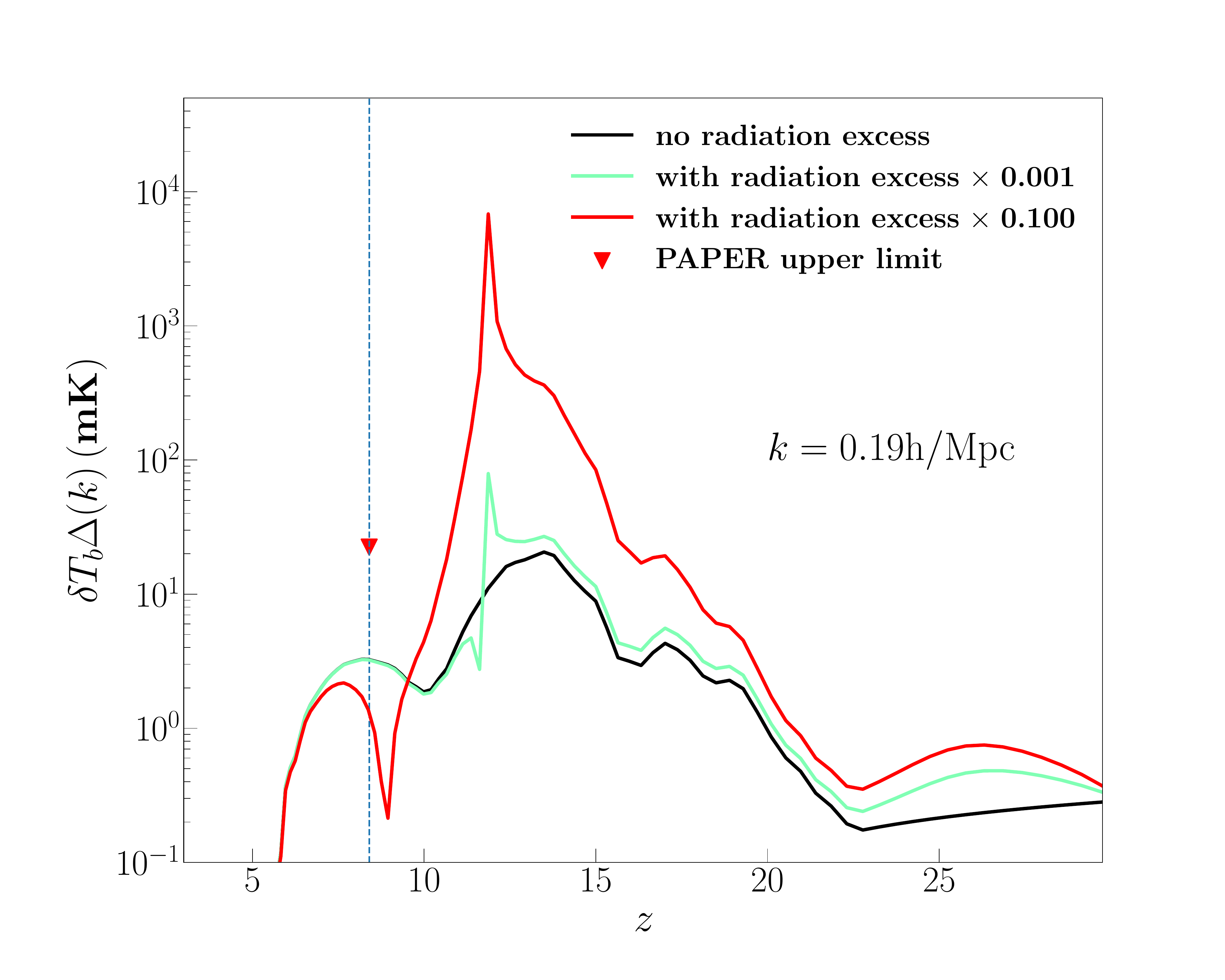}
\caption{The redshift evolution of brightness temperature fluctuations at $k=0.19 h/\rm{Mpc}$. The power of the fluctuations is defined as $\Delta^2(k)=k^3/(2\pi)P(k)$ and $P(k)$ is a 3D power spectrum of the brightness temperature fluctuations.}\label{fluc}
\end{figure}

\section{Results and discussions}

The spin temperature with 10\% of the excess is shown in Fig.~\ref{gs_standard}. We obtain the modified 21 cm global signal according to ARCADE 2 excess in Fig.~\ref{gs}. A deep trough with a temperature $\sim1.1\,\rm{K}$ can be detected around $\nu=90\,\rm{MHz}$ (Fig.~\ref{gs} (bottom)). It is apparent that if 10\% of the radio background observed today is in place at early times it will have a large effect on the \gs. Furthermore, even 0.1\% of today's background would be enough to produce an observable effect. 

In Fig.~\ref{gs} (top), it is seen that the strong coupling between the radiation field and the spin temperature remains unperturbed in the dark ages but the first trough associated with the formation of first galaxies is altered significantly. The excess Lyman-$\alpha$ heating due to the excess radiation can shift the peak of the trough to a lower redshift but the enlarged difference between the radiation field and the spin temperature due to stronger heating gives rise to a deeper absorption trough with an amplitude $\sim7$ times higher (assuming 10\% excess). 
As a consequence, the reionization due to ultraviolet radiation is also delayed because of the substantially higher radiation temperature of the 21 cm transition. 

Due to instrumental frequency coverage and sensitivity, the absorption trough may be the easiest and the most important feature to be detected from the \gs. With the enhanced absorption features in the \gs, the detection of the \gs\ would be made significantly easier.

From a 19-minute effective measurements of the LEDA experiment, a -$890\,\rm{mK}$ lower limit on the amplitude of the absorption trough is placed between 50 and 100 \rm{MHz} (Fig.~\ref{gs} (bottom)) \citep{2016MNRAS.461.2847B}. From observations within a few hours, the EDGES experiment found that the residual spectrum, which is foreground subtracted, has a $17\,\rm{mK}$ r.m.s. over 90--190 \rm{MHz} \citep{2017ApJ...847...64M}. Both limits require the excess radio fraction of high-redshift origin to be very small. 

The spectral information of the sky-averaged signal can also be drawn from the brightness temperature fluctuations at certain angular scales. The black curve in Fig.~\ref{fluc} is taken from \pc\ simulations \citep{2008ApJ...689....1S} and is the fluctuation power spectrum with no radiation excess. Based on this power spectrum and the scaling relations revealed by Fig.~\ref{gs_standard}, we can further derive two additional power spectra with 0.1\% and 10\% radiation excess. The predicted strengths of the fluctuation power at $z=8.4$ are well below the recent limit from the Precision Array for Probing the Epoch of Reionization (PAPER), leaving a broad parameter space for the excess fraction \citep{2015ApJ...809...61A}.

In this work, we have investigated the impact of an excess radiation field on the \gs, which is found to be significantly amplified, especially in absorption. We find that this key feature can be an order of magnitude stronger than expected, and would be easily detected. Future experiments, such as the DARE and the EDGES \citep{2010Natur.468..796B}, will be able to examine this enhanced global signal.\\

This research is supported by the Brand and Monica Fortner Chair. We had helpful discussions with Jonathan Sievers, Nick Gnedin, and the participants in the Radio Synchrotron Background Workshop in Richmond, VA.\\

\bibliography{21global}

\acknowledgments

\end{document}